\documentclass[reprint,onecolumn,amsmath,amssymb,jap,aip]{revtex4-1}
\usepackage{graphicx}% Include figure files
\usepackage{dcolumn}% Align table columns on decimal point
\usepackage{bm}% bold math
\usepackage{mathptmx}
\usepackage{times}
\usepackage{amsmath}
\usepackage{amssymb}
\usepackage{siunitx}
\usepackage{xspace}
\usepackage{units}

\begin{document}
\title{Temporal correlation detection using computational phase-change memory}
\author{Abu~Sebastian}\email{ase@zurich.ibm.com}
\affiliation{IBM Research -- Zurich, S\"{a}umerstrasse 4, 8803 R\"{u}schlikon, Switzerland.}
\author{Tomas Tuma}
\affiliation{IBM Research -- Zurich, S\"{a}umerstrasse 4, 8803 R\"{u}schlikon, Switzerland.}
\author{Nikolaos Papandreou}
\affiliation{IBM Research -- Zurich, S\"{a}umerstrasse 4, 8803 R\"{u}schlikon, Switzerland.}
\author{Manuel Le Gallo}
\affiliation{IBM Research -- Zurich, S\"{a}umerstrasse 4, 8803 R\"{u}schlikon, Switzerland.}
\author{Lukas Kull}
\affiliation{IBM Research -- Zurich, S\"{a}umerstrasse 4, 8803 R\"{u}schlikon, Switzerland.}
\author{Thomas Parnell}
\affiliation{IBM Research -- Zurich, S\"{a}umerstrasse 4, 8803 R\"{u}schlikon, Switzerland.}
\author{Evangelos~Eleftheriou}
\affiliation{IBM Research -- Zurich, S\"{a}umerstrasse 4, 8803 R\"{u}schlikon, Switzerland.}
%\date{\today}

\begin{abstract}
For decades, conventional computers based on the von Neumann architecture have performed computation by repeatedly transferring data between their
processing and their memory units, which are physically separated. As computation becomes increasingly data-centric and as the scalability limits in
terms of performance and power are being reached, alternative computing paradigms are searched for in which computation and storage are collocated. A
fascinating new approach is that of computational memory where the physics of nanoscale memory devices are used to perform certain computational
tasks within the memory unit in a non-von Neumann manner. Here we present a large-scale experimental demonstration using one million phase-change
memory devices organized to perform a high-level computational primitive by exploiting the crystallization dynamics. Also presented is an application
of such a computational memory to process real-world data-sets. The results show that this co-existence of computation and storage at the nanometer
scale could be the enabler for new, ultra-dense, low power, and massively parallel computing systems.
\end{abstract}
\maketitle

\begin{figure*}[h!]
\centering
\begin{tabular}{c}
\includegraphics[width = 0.7\columnwidth]{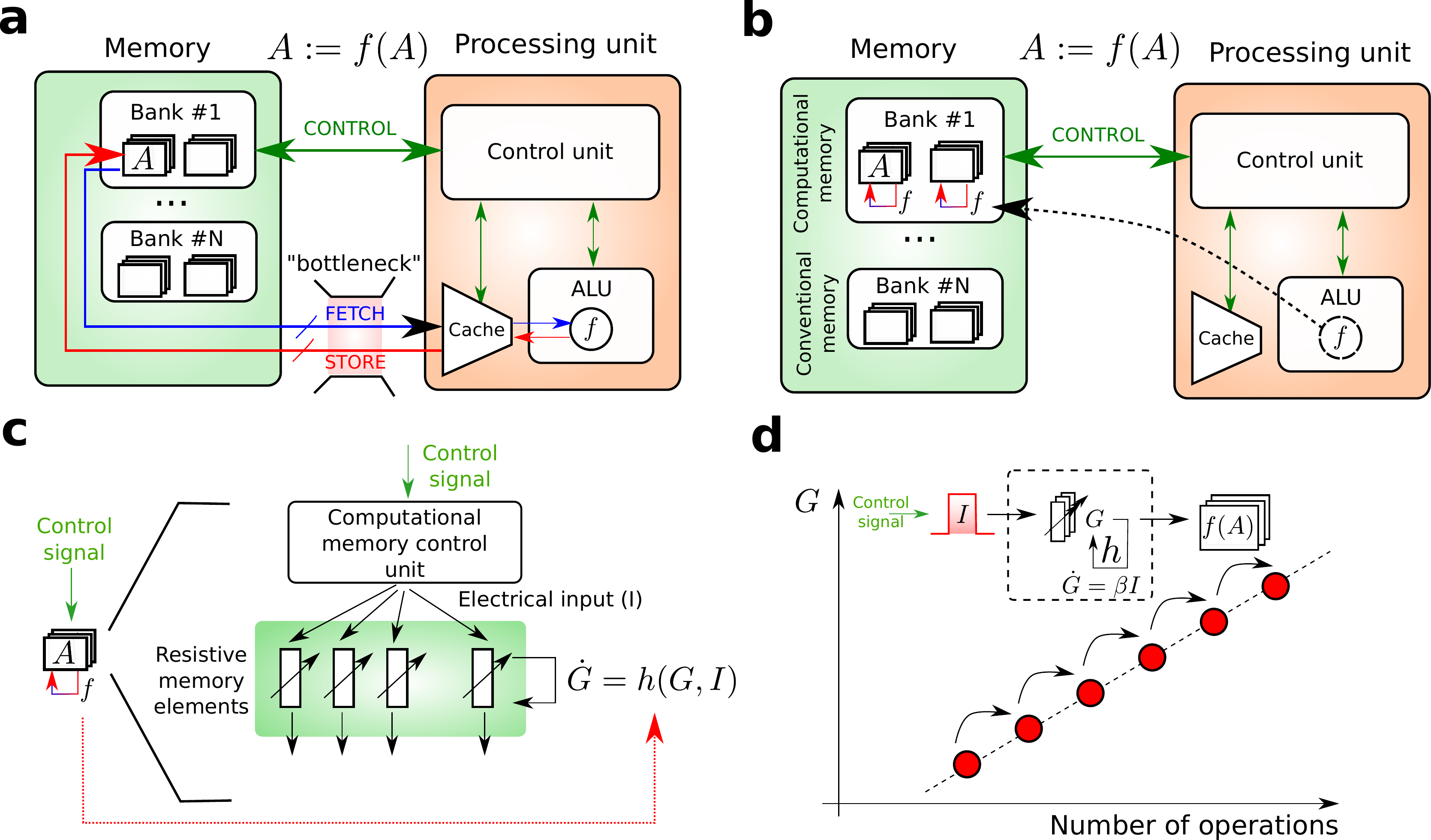}
\end{tabular}
\caption{\textbf{The concept of computational memory} (\textbf{a}) Schematic of the von Neumann computer architecture, where the memory and computing
units are physically separated. $A$ denotes information stored in a memory location. To perform a computational operation $f(A)$ and to store the
result in the same memory location, data is shuttled back and forth between the memory and the processing unit. (\textbf{b}) An alternative
architecture where $f(A)$ is performed in place in the same memory location. (\textbf{c}) Computational memory achieves this by relying on the state
dynamics of a large collection of memristive devices denoted by $h$. Depending on the operation to be performed, a suitable electrical signal is
applied to the memory devices. The conductance of the devices evolves in accordance with the electrical input, and the result of the operation can be
retrieved by reading the conductance at an appropriate time instance. (\textbf{d}) An example of the widely observed accumulative dynamics in
memristive devices. By repeated application of an electrical pulse to the devices, the conductance increases progressively.} \label{fig:intro}
\end{figure*}

In today's computing systems based on the conventional von Neumann architecture (Fig. \ref{fig:intro}(\textbf{a})), there are distinct memory and
processing units. The processing unit comprises the arithmetic and logic unit (ALU), a control unit and a limited amount of cache memory. The memory
unit typically comprises dynamic random-access memory (DRAM), where information is stored in the charge state of a capacitor. Performing an operation
(such as an arithmetic or logic operation), $f$, over a set of data stored in the DRAM, $A$, to obtain the result, $f(A)$, requires a sequence of
steps in which the data must be obtained from the memory, transferred to the processing unit, processed, and stored back to the memory. This results
in a significant amount of data being moved back and forth between the physically separated memory and processing units. This costs time and energy,
and constitutes an inherent bottleneck in performance.

To overcome this, a tantalizing prospect is that of transitioning to a hybrid architecture where certain operations, such as $f$, can be performed at
the \textit{same physical location} as where the data is stored (Fig. \ref{fig:intro}(\textbf{b})). Such a memory unit that facilitates such
collocated computation is referred to as computational memory. There are recent reports of the use of DRAM to perform bulk bit-wise operations
\cite{Y2016seshadriArXiV} and fast row copying \cite{Y2013seshadriISM} within the DRAM chip. However, a new class of emerging nanocale devices,
namely, resistive memory or memristive devices with their multi-level and non-volatile storage capability as well as sufficient richness of dynamics
are particularly well suited for computational memory. In these devices, information is stored in their resistance/conductance states and these
devices can remember the history of the current that has flowed through them \cite{Y2008strukovNature,Y2011chuaAPA,Y2015wongNatureNano}.

The essential idea behind computational memory is not to treat memory as a passive storage entity, but to exploit the dynamics of the memory devices
to realize computation exactly at the place where the data is stored. The data is stored in the conductance state of the device, denoted by $G$, as
in a conventional memory unit, and the computational task (e.g.\ $f(A)$) is performed by applying appropriate electrical signals that ``imprint" the
result of the computation in the very same memory device (Fig. \ref{fig:intro}(\textbf{c})). For example, to implement additive arithmetics,
accumulative dynamics can be used whereby the conductance evolves monotonically as a function of the number of electrical pulses applied to the
devices (Fig. \ref{fig:intro}(\textbf{d})). Besides its collocated nature, this type of computation is massively parallel, and achieves high
areal/power efficiency as well as speed.

An early proposal for the use of memristive devices for in-place computing was the realization of certain logical operations using a circuit based on
TiO$_x$-based memory devices \cite{Y2010borghettiNature}. The same memory devices are used simultaneously to store the inputs, perform the logic
operation, and store the resulting output. More complex logic units based on this initial concept were proposed
\cite{Y2014kvatinskyTCAS,Y2016zhaICCAD,Y2016vourkasIEEECSM}. Implementations of logical operations using other physical phenomena were also proposed,
such as ferroelectric domain switching induced by the tip of a scanning probe microscope \cite{Y2014ievlevNatPhys} and the
crystallization\cite{Y2013ielminiAM} and melting \cite{Y2014lokePNAS} of phase-change materials. In these memristive logic applications, one exploits
only the binary resistance states of the devices. Matrix vector multiplication using Ohm's law and Kirchhoff's law is another computational primitive
that can be realized very efficiently using resistive memory devices. Massively parallel, memory-centric hardware accelerators based on this concept
are now becoming an important subject of research \cite{Y2015burrITED,Y2016shafieeISCA,Y2016chiISCA,Y2016bojnordiHPCA,Y2017burrAPX}. However, to
date, a few experimental reports go beyond a proof-of-concept demonstration.

In this paper, we present an algorithm whereby the crystallization dynamics of a type of resistive memory device, namely, phase-change memory (PCM)
devices, are exploited to implement a high-level computational primitive. Subsequently, we present a large-scale experimental demonstration of the
computational memory concept using an array of one million PCM devices. We also present an application of such a computational memory to process
real-world data-sets such as weather data.

\section{Results}
\subsection*{Dynamics of phase-change memory devices}

\begin{figure*}[h!]
\centering
\begin{tabular}{c}
\includegraphics[width = 0.7\columnwidth]{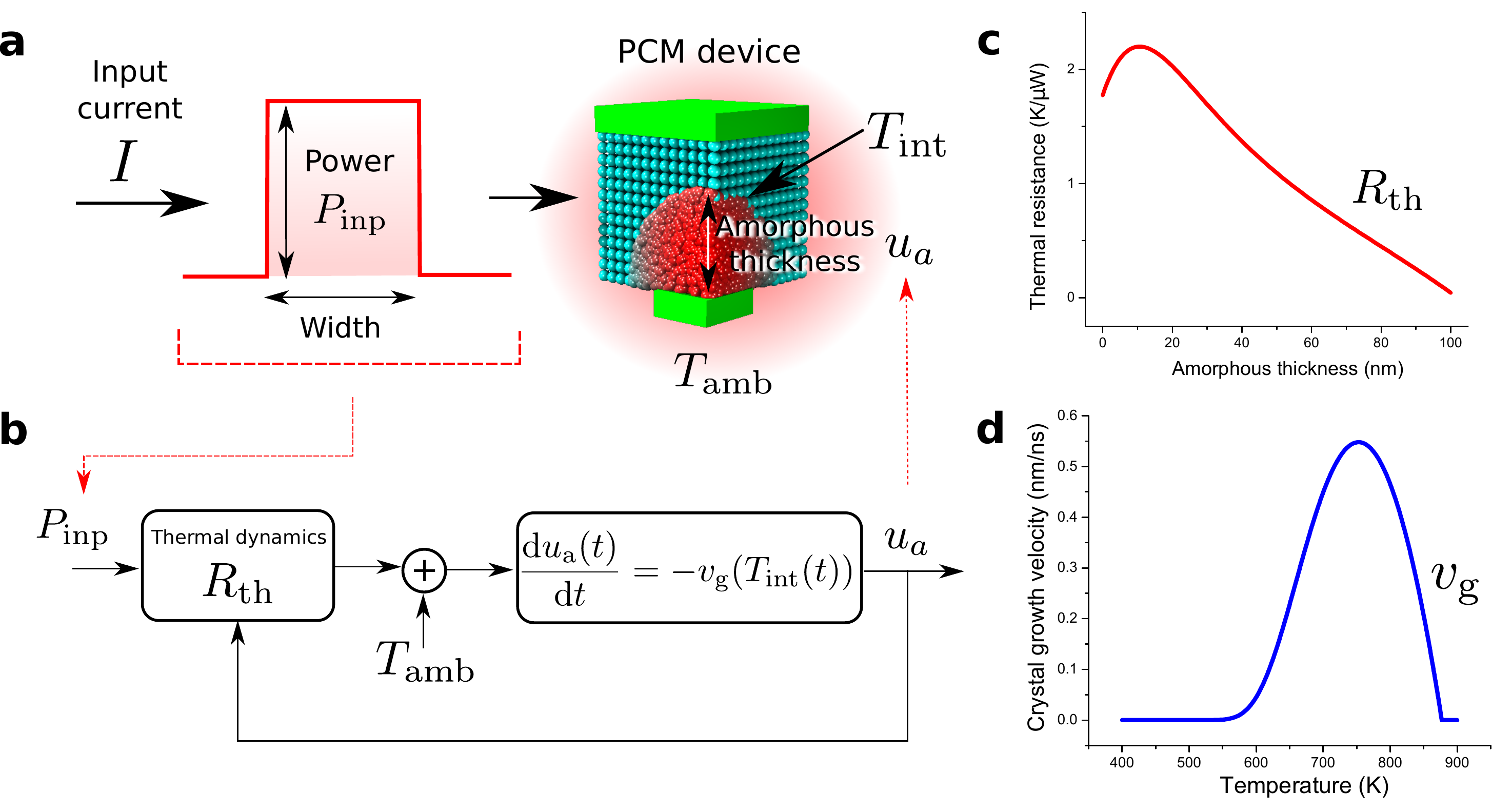}
\end{tabular}
\caption{\textbf{Crystallization dynamics} (\textbf{a}) Schematic of a mushroom-type PCM device showing the phase configurations. (\textbf{b})
Illustration of the crystallization dynamics. When an electrical signal with power $P_\text{inp}$ is applied to a PCM device, significant Joule
heating occurs. The resulting temperature distribution across the device is determined by the thermal environment, in particular the effective
thermal resistance, $R_\text{th}$. The effective thickness of the amorphous region, $u_a$, evolves in accordance with the temperature at the
amorphous--crystalline interface, $T_\text{int}$, and the temperature dependence of crystal growth, $v_\text{g}$. Experimental estimates of
(\textbf{c}) $R_\text{th}$ and (\textbf{d}) $v_\text{g}$.} \label{fig:pcmdyn}
\end{figure*}

A PCM device consists of a nanometric volume of phase-change material sandwiched between two electrodes. A schematic illustration of a PCM device
with mushroom-type device geometry is shown in Fig. \ref{fig:pcmdyn}(\textbf{a})) \cite{Y2016burrJETCAS}. In an as-fabricated device, the material is
in the crystalline phase. When a current pulse of sufficiently high amplitude is applied to the PCM device (typically referred to as the RESET
pulse), a significant portion of the phase-change material melts due to Joule heating. When the pulse is stopped abruptly, the molten material
quenches into the amorphous phase because of glass transition. In the resulting RESET state, the device will be in the low conductance state as the
amorphous region blocks the bottom electrode. The size of the amorphous region is captured by the notion of an effective thickness, $u_a$
\cite{Y2011sebastianJAP}. PCM devices exhibit a rich dynamic behavior with an interplay of electrical, thermal and structural dynamics that forms the
basis for their application as computational memory. The electrical transport exhibits a strong field and temperature dependence
\cite{Y2015legalloNJP}. Joule heating and the thermal transport pathways ensure that there is a strong temperature gradient within the PCM device.
Depending on the temperature in the cell, the phase-change material undergoes structural changes, such as phase transitions and structural relaxation
\cite{Y2014sebastianNatComm,Y2015ratyNatComm}.

In our demonstration, we focus on a specific aspect of the PCM dynamics: the crystallization dynamics capturing the progressive reduction in the size
of the amorphous region due to the amorphous to crystalline phase transition (Fig. \ref{fig:pcmdyn}(\textbf{b})). When a current pulse (typically
referred to as the SET pulse) is applied to a PCM device in the RESET state such that the temperature reached in the cell via Joule heating is high
enough, but below the melting temperature, a part of the amorphous region crystallizes. At the nanometer scale, the crystallization mechanism is
dominated by crystal growth due to the large amorphous--crystalline interface area and the small volume of the amorphous region
\cite{Y2014sebastianNatComm}. The crystallization dynamics in such a PCM device can be approximately described by
\begin{equation}\label{eqn:cryst}
\frac{\text{d}u_\text{a}}{\text{d}t} = -v_\text{g}(T_\text{int}),
\end{equation}
where $v_\text{g}$ denotes the temperature-dependent growth velocity of the phase-change material; $T_\text{int} =
R_\text{th}(u_\text{a})P_\text{inp} + T_\text{amb}$ is the temperature at the amorphous--crystalline interface, and $u_\text{a}(0) = u_{\text{a}_0}$
is the initial effective amorphous thickness \cite{Y2014sebastianNatComm}. $T_\text{amb}$ is the ambient temperature, and $R_\text{th}$ is the
effective thermal resistance that captures the thermal resistance of all possible heat pathways. Experimental estimates of $R_\text{th}$ and
$v_\text{g}$ are shown in Fig. \ref{fig:pcmdyn}(\textbf{c}) and Fig. \ref{fig:pcmdyn}(\textbf{d}), respectively \cite{Y2014sebastianNatComm}. From
the estimate of $R_\text{th}$ as a function of $u_\text{a}$, one can infer that the hottest region of the device is slightly above the bottom
electrode and that the temperature within the device monotonically decreases with increasing distance from the bottom electrode. The estimate of
$v_\text{g}$ shows the strong temperature dependence of the crystal growth rate. Up to approx. \unit[550]{K}, the crystal growth rate is negligible
whereas it is maximum at approx. \unit[750]{K}. As a consequence of Equation \eqref{eqn:cryst}, $u_\text{a}$ progressively decreases upon the
application of repetitive SET pulses and hence the low-field conductance progressively increases. In subsequent discussions, the RESET and SET pulses
will be collectively referred to as write pulses. It is also worth noting that in a circuit-theoretic representation, the PCM device can be viewed as
a generic memristor, with $u_a$ serving as an internal state variable \cite{Y2015corintoJETCAS,Y2016ascoliIJCTA}.

\subsection*{Detecting statistical correlations using computational memory}
\begin{figure*}[t!]
\centering
\begin{tabular}{c}
\includegraphics[width = 0.7\columnwidth]{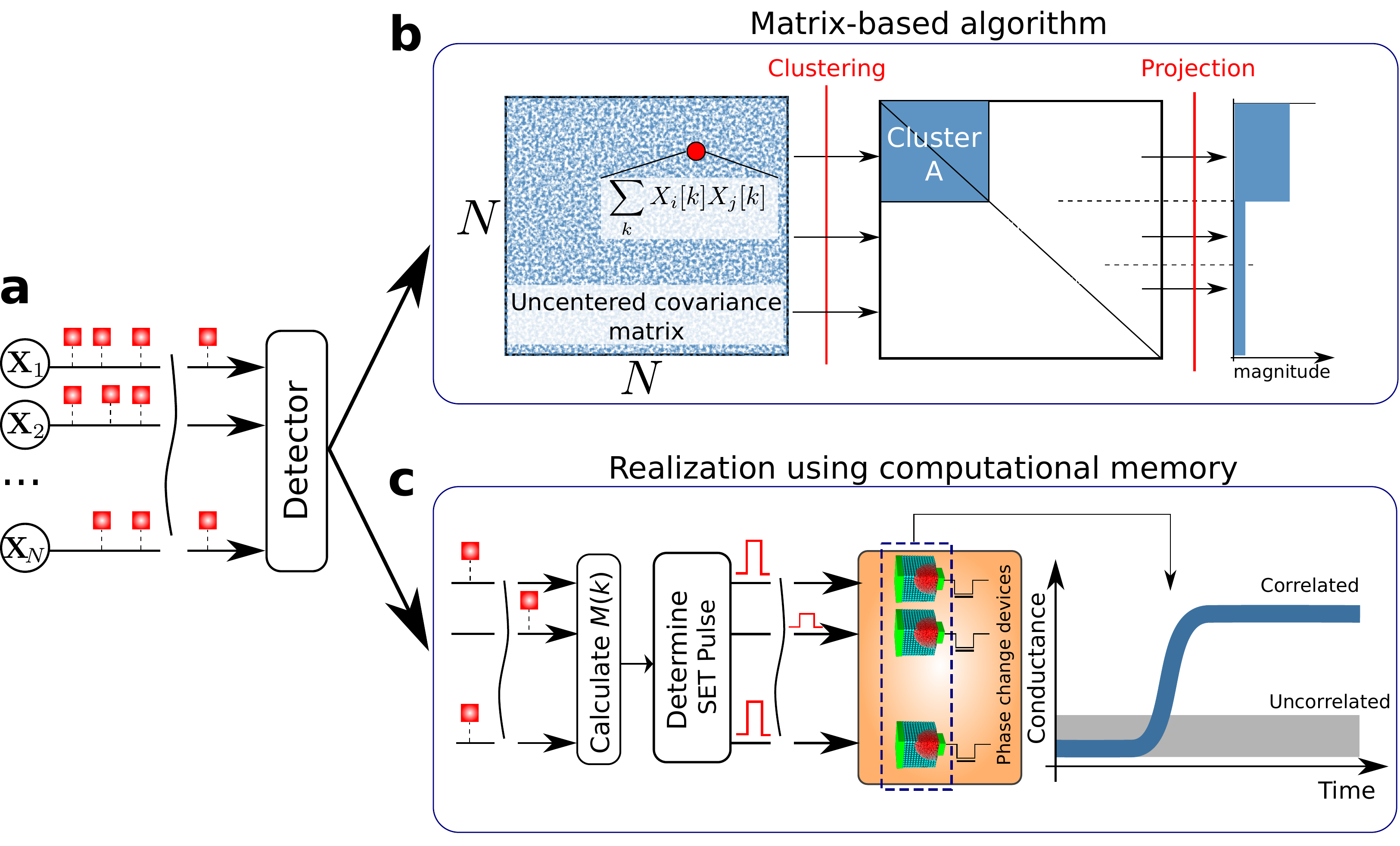}
\end{tabular}
\caption{\textbf{Temporal correlation detection} (\textbf{a}) Schematic of $N$ stochastic binary processes, some correlated and the remainder
uncorrelated, arriving at a correlation detector. (\textbf{b}) One approach to detect the correlated groups is to obtain an uncentered covariance
matrix. By summing the elements of this matrix along a row or column, we can obtain some kind of numerical weights corresponding to the $N$ processes
and can differentiate the correlated from the uncorrelated group based on their magnitude. (\textbf{c}) Alternatively, the correlation detection
problem can be realized using computational memory. Here, each process is assigned to a single PCM device. Whenever the process takes the value 1, a
SET pulse is applied to the PCM device. The amplitude or the width of the SET pulse is chosen to be proportional to the instantaneous sum of all
processes. By monitoring the conductance of the memory devices, we can decipher the correlated groups.} \label{fig:alg}
\end{figure*}

In this section, we show how the crystallization dynamics of PCM devices can be exploited to perform a high-level computational task, such as that of
detecting statistical correlations between event-based data streams. Such problems arise in a multitude of application areas such as, the Internet of
Things (IoT), life sciences, networking, social networks and large scientific experiments \cite{Y2014lazerScience}. For example, one could generate
an event-based data stream based on the presence or absence of a specific word in a collection of tweets. Sensory data processing is another
promising application area, especially with sensors such as the dynamic vision sensor \cite{Y2010liuCON}. One can also view correlation detection as
a key constituent of unsupervised learning where one of the objectives is to find correlated clusters in data streams.

In a generic formulation of the problem, let us assume that there are $N$ discrete-time binary stochastic process arriving at a correlation detector
(see Fig. \ref{fig:alg}(\textbf{a})). Let $\textbf{X}_i = \{X_i(k)\}$ be one of the processes. Then $X_i(k)$ is a random variable with probabilities
\begin{eqnarray}\label{eqn:prob1}
P[X_i(k) = 1] &=& p\\\label{eqn:prob2} P[X_i(k)=0] &=& 1-p,
\end{eqnarray}
for $0\leq p \leq 0.5$. Let $\textbf{X}_j$ be another discrete-time binary stochastic processes with the same value of parameter $p$. Then the
correlation coefficient of the random variables $X_i(k)$ and $X_j(k)$ at time instant $k$ is defined as
\begin{eqnarray}\label{eqn:c}
c &=& \frac{Cov[X_i(k),X_j(k)]}{\sqrt{Var[X_i(k)]Var[X_j(k)]}}.
\end{eqnarray}
Processes $\textbf{X}_i$ and $\textbf{X}_j$ are said to be correlated if $c>0$ or uncorrelated otherwise. The objective of the correlation detection
problem is to detect, in an unsupervised manner, an unknown subset of these processes that are mutually correlated.

As schematically illustrated in Fig. \ref{fig:alg}(\textbf{b}), one way to solve this problem is by obtaining an estimate of the uncentered
covariance matrix corresponding the processes denoted by
\begin{equation}\label{eqn:Rijest}
\hat{R}_{ij} = \frac{1}{K}\sum_{k=1}^KX_i(k)X_j(k).
\end{equation}
Next, by summing the elements of this matrix along a row or column, we can obtain certain numerical weights corresponding to the processes denoted by
$\hat{W}_i = \sum_{j=1}^{N} \hat{R}_{ij}$. It can be shown that if $\textbf{X}_i$ belongs to the correlated group with correlation coefficient $c>0$,
then
\begin{equation}\label{eqn:wc}
E[\hat{W}_i] = (N-1)p^2 + p + (N_\text{c}-1)c p(1-p).
\end{equation}
$N_\text{c}$ denotes the number of processes in the correlated group. In contrast, if $\textbf{X}_i$ belongs to the uncorrelated group, then
\begin{equation}\label{eqn:wuc}
E[\hat{W}_i] = (N-1)p^2 + p.
\end{equation}
Hence by monitoring $\hat{W}_i$ in the limit of large $K$, we can determine which processes are correlated with $c>0$. Moreover, it can be seen that
with increasing $c$ and $N_\text{c}$, it becomes easier to determine whether a process belongs to a correlated group.

We can show that this rather sophisticated problem of correlation detection can be efficiently solved using a computational memory module comprising
PCM devices by exploiting the crystallization dynamics. By assigning each incoming process to a single PCM device, the statistical correlation can be
calculated and stored in the very same device as the data passes through the memory. The way it is achieved is schematically depicted in Fig.
\ref{fig:alg}(\textbf{c}): At each time instance, $k$, a ``collective momentum", $M(k) = \sum_{j=1}^N X_j(k)$, that corresponds to the instantaneous
sum of all processes is calculated. The calculation of $M(k)$ incurs little computational effort as it just counts the number of non-zero events at
each time instance. Next, an identical SET pulse is applied potentially in parallel to all the PCM devices for which the assigned binary process has
a value of 1. The duration or amplitude of the SET pulse is chosen to be a linear function of $M(k)$. For example, let the duration of the pulse,
$\delta t(k) = CM(k) = C\sum_{j=1}^N X_j(k)$. For the sake of simplicity, let us assume that the interface temperature, $T_\text{int}$, is
independent of the amorphous thickness, $u_\text{a}$. As the pulse amplitude is kept constant, $v_\text{g}(T_\text{int}) = \mathcal{G}$, where
$\mathcal{G}$ is a constant. Then from Equation \ref{eqn:cryst}, the absolute value of the change in the amorphous thickness of the $i^\text{th}$
phase-change device at the $k^\text{th}$ discrete-time instance is
\begin{equation}\label{eqn:deltaua}
\delta u_{\text{a}_i}(k) = \delta t(k)v_\text{g}(T_\text{int}) = C\mathcal{G}\sum_{j=1}^N X_j(k)
\end{equation}
The total change in the amorphous thickness after $K$ time steps can be shown to be
\begin{eqnarray}\nonumber
\Delta u_{\text{a}_i}(K) &=& \sum_{k=1}^K \delta u_{\text{a}_i}(k) X_i(k)\\\nonumber &=&C\mathcal{G} \sum_{k=1}^K\sum_{j=1}^N X_i(k)X_j(k)\\\nonumber
&=& C\mathcal{G} \sum_{j=1}^N\sum_{k=1}^K X_i(k)X_j(k)\\\nonumber &=& KC\mathcal{G}\sum_{j=1}^N\hat{R}_{ij}\\\label{eqn:Deltaua} &=&
KC\mathcal{G}\hat{W}_i.
\end{eqnarray}

Hence, from Equations \ref{eqn:wc} and \ref{eqn:wuc}, if $\textbf{X}_i$ is one of the correlated processes, then $\Delta u_{\text{a}_i}$ will be
larger than if $\textbf{X}_i$ is one of the uncorrelated processes. Therefore by monitoring $\Delta u_{\text{a}_i}$ or the corresponding
resistance/conductance for all phase-change devices, we can determine which processes are correlated.

\subsection*{Experimental platform}

\begin{figure*}[h!]
\centering
\begin{tabular}{c}
\includegraphics[width = 0.7\columnwidth]{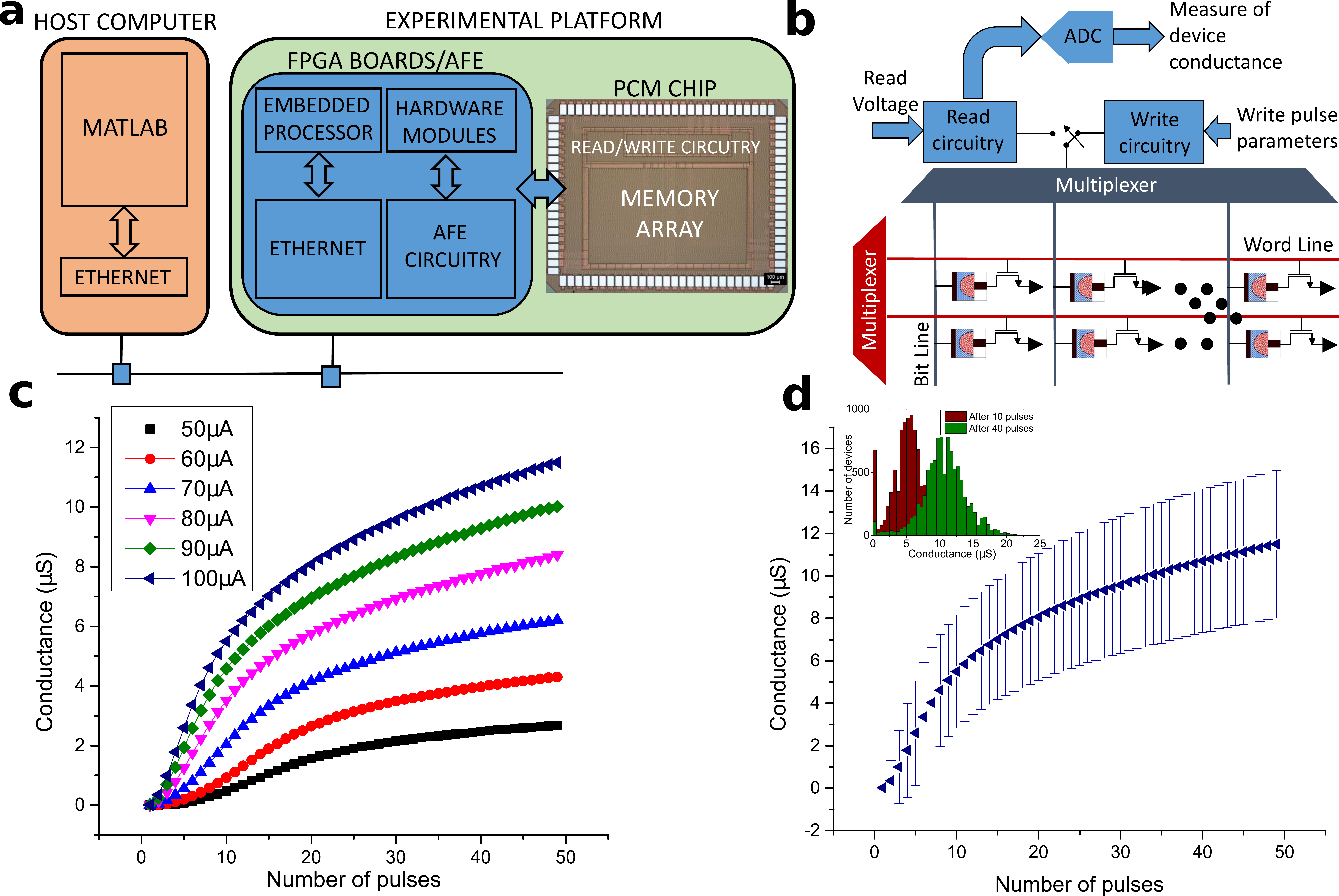}
\end{tabular}
\caption{\textbf{Experimental platform and characterization results} (\textbf{a}) Schematic illustration of the experimental platform showing the
main components. (\textbf{b}) The PCM array is organized as a matrix of word lines (WL) and bit lines (BL), and the chip also integrates the
associated read/write circuitries. (\textbf{c}) The mean accumulation curve of 10,000 devices showing the map between the device conductance and the
number of pulses. The devices achieve a higher conductance value with increasing SET current and also with increasing number of pulses. (\textbf{d})
The mean and standard deviation associated with the accumulation curve corresponding to the SET current of \unit[100]{$\mu$A}. Also shown are the
distributions of conductance values obtained after application of the 10th and 40th SET pulses.} \label{fig:expplat}
\end{figure*}

Next, we present experimental demonstrations of the concept. The experimental platform (schematically shown in Fig. \ref{fig:expplat}(\textbf{a})) is
built around a prototype PCM chip that comprises 3 million PCM devices. More details on the chip are presented in the methods section. As shown in
Fig. \ref{fig:expplat}(\textbf{b})), the PCM array is organized as a matrix of word lines (WL) and bit lines (BL). In addition to the PCM devices,
the prototype chip integrates the circuitry for device addressing and for write and read operations. The PCM chip is interfaced to a hardware
platform comprising two field programmable gate array (FPGA) boards and an analog-front-end (AFE) board. The AFE board provides the power supplies as
well as the voltage and current reference sources for the PCM chip. The FPGA boards are used to implement the overall system control and data
management as well as the interface with the data processing unit. The experimental platform is operated from a host computer, and a Matlab
environment is used to coordinate the experiments.

An extensive array-level characterization of the PCM devices was conducted prior to the experimental demonstrations. In one experiment, 10,000
devices were arbitrarily chosen and were first RESET by applying a rectangular current pulse of \unit[1]{$\mu$s} duration and \unit[440]{$\mu$A}
amplitude. After RESET, a sequence of SET pulses of \unit[50]{ns} duration were applied to all the devices, and the resulting device conductance
values were monitored after the application of each pulse. The map between the device conductance and the number of pulses is referred to as an
accumulation curve. The accumulation curves corresponding to different SET currents are shown in Fig. \ref{fig:expplat}(\textbf{c}). These results
clearly show that the mean conductance increases monotonically with increasing SET current (in the range from \unit[50]{$\mu$A} and
\unit[100]{$\mu$A}) and with increasing number of SET pulses. From Fig. \ref{fig:expplat}(\textbf{d}), it can also be seen that there is significant
variability associated with the evolution of the device conductance values. This variability arises from inter-device as well as intra-device
variability. The intra-device variability is traced to the differences in the atomic configurations of the amorphous phase created via the
melt-quench process after each RESET operation \cite{Y2016tumaNatureNano,Y2016legalloESSDERC}. Besides the variability arising from the
crystallization process, additional fluctuations in conductance also arise from $1/f$ noise \cite{Y2010closeIEDM} and drift variability
\cite{Y2012pozidisIMW}.

\subsection*{Experimental demonstration with a million correlated processes}

\begin{figure*}[h!]
\centering
\begin{tabular}{c}
\includegraphics[width = 0.9\columnwidth]{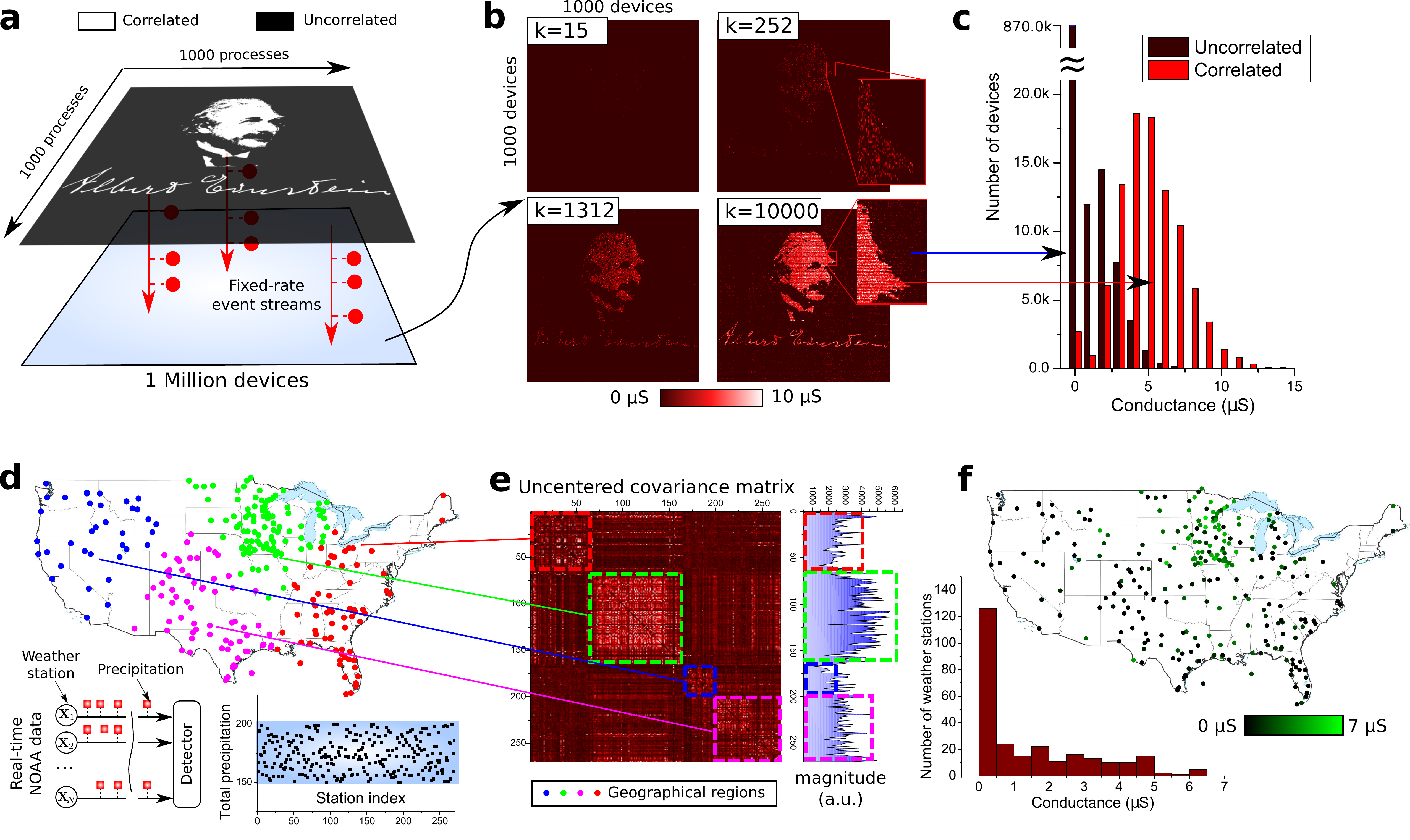}
\end{tabular}
\caption{\textbf{Experimental results} (\textbf{a}) A million processes are mapped to the pixels of a 1000 $\times$ 1000 pixel black-and-white
composite image of Albert Einstein and his signature [both taken from \cite{wikipedia}]. The pixels turn on and off in accordance with the
instantaneous binary values of the processes. (\textbf{b}) Evolution of device conductance over time, showing that the devices corresponding to the
correlated processes go to a high conductance state. (\textbf{c}) Distribution of the device conductance shows that the algorithm is able to pick out
most of the correlated processes. (\textbf{d}) Generation of a binary stochastic process based on the rainfall data from 270 weather stations across
the USA. (\textbf{e}) The uncentered covariance matrix reveals several small correlated groups, along with a predominant correlated group.
(\textbf{f}) Map of the device conductance levels after the experiment shows that the devices corresponding to the predominant correlated group have
achieved a higher conductance value.} \label{fig:4}
\end{figure*}

In a first demonstration of correlation detection, we created the input data artificially, and generated one million binary stochastic processes
organized in a two-dimensional grid (Fig. \ref{fig:4}(\textbf{a})). We arbitrarily chose a subset of 95,525 processes, which we mutually correlated
with a relatively weak instantaneous correlation coefficient of 0.1, whereas the other 904,475 were uncorrelated. The objective was to see if we can
detect these correlated processes using the computational memory approach. Each stochastic process was assigned to a single PCM device. First, all
devices were RESET by applying a current pulse of \unit[1]{$\mu$s} duration and \unit[440]{$\mu$A} amplitude. In this experiment, we chose to
modulate the SET current while maintaining a constant pulse duration of \unit[50]{ns}. At each time instance, the SET current is chosen to be equal
to $0.002*M(k)$ $\mu$A, where $M(k) = \sum_{j=1}^N X_j(k)$ is equal to the collective momentum. This rather simple calculation was performed in the
host computer. Alternatively, it could be done in one of the FPGA boards. Next, the on-chip write circuitry was instructed to apply a SET pulse with
the calculated SET current to all the PCM devices for which $X_i(k)=1$. To minimize the execution time, we chose not to program the devices if the
SET current is less than \unit[25]{$\mu$A}. The SET pulses were applied sequentially. However, if the chip has multiple write circuitries that can
operate in parallel, then it is also possible to apply the SET pulses in parallel. This process of applying SET pulses was repeated at every time
instance. The maximum SET current applied to the devices during the experiment was \unit[80]{$\mu$A}.

As described earlier, owing to the temporal correlation between the processes, the devices assigned to those processes are expected to go to a high
conductance state. We periodically read the conductance values of all PCM devices using the on-chip read circuitry and the on-chip ADC. The resulting
map of the conductance values are shown in Fig. \ref{fig:4}(\textbf{b}). Also shown is the corresponding distribution of the conductance values (Fig.
\ref{fig:4}(\textbf{c})). This distribution shows that we can distinguish between the correlated and the uncorrelated processes. The inaccuracies are
attributed to the variability and conductance fluctuations discussed earlier. However, it is remarkable that in spite of these issues, we are able to
perform the correlation detection with significantly high accuracy. Note that there are several applications, such as sensory data processing, where
these levels of accuracy could be sufficient. Moreover, there are also ways to improve the accuracy by using multiple devices to interface with a
single random process. The conductance fluctuations can also be minimized using concepts such as projected phase change memory
\cite{Y2015koelmansNatComm}.

Another issue is that of the limited dynamic range of the conductance values. There is a limit to the $u_\text{a}$ and hence the maximum conductance
values that the devices can achieve. The accumulation curves in Fig. \ref{fig:expplat}(\textbf{d}) clearly show that the mean conductance values
begin to saturate after the application of a certain number of pulses. This means that the correlations have to be detected within a reasonable
amount of time dictated by this ``dynamic range". Hence, once the correlation has been detected, then the devices need to be RESET, and the operation
has to be resumed to detect subsequent correlations. The application of shorter SET pulses is one way to increase the dynamic range. Unfortunately,
in the PCM chip we experimented with, we were limited to a minimum pulse width of \unit[50]{ns}. The use of multiple devices to interface with the
random processes can also enlarge the dynamic range.

As per Equation (\ref{eqn:wc}), we would expect the level of separation between the distributions of correlated and uncorrelated groups to increase
with increasing values of the correlation coefficient. This could be experimentally confirmed. The experimental results show that the correlated
groups can be detected down to very low correlation coefficients such as $c=0.01$ even though it will be difficult to make a precise evaluation of
the correlation coefficient. Moreover, this concept could potentially also be used to detect multiple correlated groups having different correlation
coefficients.

\subsection*{Experiment with the weather data}
A second demonstration is based on real-world data from 270 weather stations across the USA. Over a period of 6 months, the rainfall data from each
station constituted a binary stochastic process that was applied to the computational memory at one-hour time steps. The process took the value 1 if
rainfall occurred in the preceding one-hour time window, else it was 0 (Fig. \ref{fig:4}(\textbf{d})). An analysis of the uncentered covariance
matrix shows that several correlated groups exist and that one of them is predominant. As expected, also a strong geographical correlation with the
rainfall data exists (Fig. \ref{fig:4}(\textbf{e})). Correlations between the rainfall events are also reflected in the geographical proximity
between the corresponding weather stations. To detect the predominant correlated group using computational memory, we used the same approach as
above, but with 4 PCM devices interfacing with each weather station data. The four devices were used to improve the accuracy. At each instance in
time, the SET current was calculated to be equal to $0.0013*M(k)$ $\mu$A. Next, the PCM chip was instructed to program the $270\times 4$ devices
sequentially with the calculated SET current. The on-chip write circuitry applies a write pulse with the calculated SET current to all PCM devices
for which $X_i(k)=1$. We chose not to program the devices if the SET current is less than \unit[25]{$\mu$A}. The duration of the pulse was fixed to
be \unit[50]{ns}, and the maximum SET current applied to the devices was \unit[80]{$\mu$A}. The resulting device conductance map (averaged over the
four devices per weather station) shows that the conductance values corresponding to the predominant correlated group of weather stations are
comparably higher (Fig. \ref{fig:4}(\textbf{f})).

Based on a threshold conductance value chosen to be \unit[2]{$\mu$S}, we can classify the weather stations into correlated and uncorrelated weather
stations. We can also make comparisons with established unsupervised classification techniques such as $k$-means clustering. It was seen that, out of
the 270 weather stations, there was a match for 245 weather stations. The computational memory approach classified 12 stations as uncorrelated which
were marked correlated by the $k$-means clustering approach. Similarly, the computational memory approach classified 13 stations as correlated which
were marked uncorrelated by the $k$-means clustering approach. Given the simplicity of the computational memory approach, it is remarkable that it
can achieve this level of similarity with such a sophisticated and well-established classification algorithm.

\section{Discussion}
The scientific relevance of the presented work is that we have convincingly demonstrated the ability of computational memory to perform certain
high-level computational tasks in a non-von Neumann manner by exploiting the dynamics of resistive memory devices. We have also demonstrated the
concept experimentally at the scale of a million PCM devices. Even though in the experimental demonstrations using the prototype chip, we programmed
the devices sequentially, we could also program them in parallel given the availability of sufficient number of write modules. A hypothetical
computational memory module performing correlation detection need not be substantially different from conventional memory modules. The main
constituents of such a module will also be a memory controller and a memory chip. Tasks such as computing $M(k)$ can easily be performed in the
memory controller. The memory controller can then convey the write/read instructions to the memory chip. It can be shown that in such a computational
memory module, we could accelerate the correlation detection task by a factor of 200 compared to state-of-the-art computing hardware. In order to
quantify the time, we measured the performance of various different implementations that can be executed on an IBM Power System S822LC system. This
system has 2 POWER8 CPUs (each comprising 10 cores) and 4 Nvidia Tesla P100 graphical processing units (attached using the NVLink interface).

An alternate approach to using PCM devices will be to design an application-specific chip where the accumulative behavior of PCM is emulated using
complementary metal-oxide semiconductor (CMOS) technology using adders and registers. However, even at a relatively large \unit[90]{nm}-technology
node, the areal footprint of the computational phase-change memory is much smaller than that of CMOS-only approaches even though the dynamic power
consumption is comparable. By scaling the devices to smaller dimensions and by using shorter write pulses, these gains are expected to increase
several fold \cite{Y2011xiongScience,Y2012lokeScience}. The ultra-fast crystallization dynamics and non-volatility ensure a multi-time-scale
operating window ranging from a few tens of nanoseconds to years. These attributes are particularly attractive for slow processes, where the leakage
of CMOS would dominate dynamic power because of low utilization rate.

It can be shown that a single-layer spiking neural network can also be used to detect temporal correlations \cite{Y2003guetigJN}. The event-based
data-streams can be translated to pre-synaptic spikes to a synaptic layer. Based on the synaptic weights, the postsynaptic potentials are generated
and added to the membrane potential of a leaky integrate-and-fire neuron. The temporal correlations between the presynaptic input spikes and the
neuronal firing events result in an evolution of the synaptic weights due to a feedback-driven competition among the synapses. In the steady state,
the correlations between the individual input streams can be inferred from the distribution of the synaptic weights or the resulting firing activity
of the postsynaptic neuron. Recently, it was shown that in such a neural network, PCM devices can serve as the synaptic elements \cite{Y2016tumaEDL}.
One could argue that the synaptic elements serve as some form of computational memory. Even though both approaches aim to solve the same problem,
there are some notable differences. In the neural network approach, it is the spike-timing-dependent-plasticity rule and the network dynamics that
enable correlation detection. One could use any passive multi-level storage element to store the synaptic weight. Also note that the neuronal input
is derived based on the value of the synaptic weights. It is challenging to implement such a feedback architecture within a computational memory
unit. Such feedback architectures are also likely to be much more sensitive to device variabilities and nonlinearities and are not well suited for
detecting very low correlations \cite{Y2003guetigJN}.

Detection of statistical correlations is just one of the computational primitives that could be realized using the crystallization dynamics. Another
application of crystallization dynamics is that of finding factors of numbers proposed originally by Wright et
al.\cite{Y2011wrightAdvMat,Y2013wrightAFM,Y2015hosseiniEDL}. Assume that a PCM device is initialized in such a way that after the application of $X$
number of pulses, the conductance exceeds a certain threshold. To check whether $X$ is a factor of $Y$, $Y$ number of pulses are applied to the
device, re-initializing the device each time the conductance exceeds the threshold. It can be seen that if after the application of $Y$ pulses, the
conductance of the device is above the threshold, then $X$ is a factor of $Y$. Another fascinating application of crystallization dynamics is to
realize matrix-vector multiplications. To multiple an $N\times N$ matrix, $A$, with a $N\times 1$vector, $x$, the elements of the matrix and the
vector can be translated into the durations and amplitudes of a sequence of crystallizing pulses applied to an array of $N$ PCM devices. It can be
shown that by monitoring the conductance levels of the PCM devices, one gets a good estimate of the matrix-vector product. Note that such an approach
is superior to the more common approach using Ohm's law and Kirchhoff's law that require $N\times N$ devices.

In addition to crystallization dynamics, one could also exploit other rich dynamic behavior in PCM devices, such as the dynamics of structural
relaxation. Whenever an amorphous state is formed via the melt-quench process, the resulting unstable glass state relaxes to an energetically more
favorable ``ideal" glass state \cite{Y2011boniardiAPL,Y2012fantiniAPL,Y2015ratyNatComm,Y2016zipoliPRB}. This structural relaxation, which codes the
temporal information of the application of write pulses, can be exploited to perform tasks such as detection of rates of processes in addition to
their temporal correlations. It is also foreseeable that by further coupling the dynamics of these devices we can potentially solve even more
intriguing problems. Suggestions of such memcomputing machines that could solve certain non-deterministic polynomial (NP) problems in polynomial (P)
time by exploiting attributes such as inherent parallelism, functional polymorphism and information overhead are being actively investigated
\cite{Y2015traversaTNNLS,Y2015diventraSciAm}. The concepts presented in this work could also be extended to the optical domain using devices such as
photonic PCM \cite{Y2015riosNaturePhotonics}. In such an approach, optical signals will be used to program the devices instead of electrical signals.
These concepts are also not limited to PCM devices: several other memristive device technologies exist that possess sufficiently rich dynamics to
serve as computational memory \cite{Y2007waserNatMat}. However, it is worth noting that PCM technology is arguably the most advanced resistive memory
technology at present with a very well established multi-level storage capability \cite{Y2016burrJETCAS}. The read endurance is assumed to be
unlimited. There are also recent reports of more than 10$^{12}$ RESET/SET endurance \cite{Y2016kimIEDM}. Note that in our experiments, we mostly
apply only the SET pulses, and in this case the endurance is expected to be substantially higher.

To summarize, we have presented the first significant experimental demonstration of the concept of computational memory where the rich dynamic
behavior of phase-change memory devices is used to execute a high-level machine-learning algorithm almost entirely in the memory array. We have
demonstrated the co-existence of computation and storage at the nanometer scale that could enable massively parallel computing systems of the future.

\section*{Methods}
\subsection*{Phase-change memory chip}
The PCM devices were integrated into the chip in \unit[90]{nm} CMOS technology \cite{Y2010closeIEDM}. The phase-change material is doped
Ge$_2$Sb$_2$Te$_2$ (d-GST). The bottom electrode has a radius of $\sim 20$ nm and a length of $\sim 65$ nm, and was defined using a sub-lithographic
key-hole transfer process \cite{Y2007breitwischVLSI}. The phase-change material is $\sim 100$ nm thick and extends to the top electrode. Two types of
devices are available on-chip. They differ by the size of their access transistor. The first sub-array contains 2 million devices. In the second
sub-array, which contains 1 million devices, the access transistors are twice as large. All experiments in this work were done on the second
sub-array, which is organized as a matrix of 512 word lines (WL) and 2048 bit lines (BL). The selection of one PCM device is done by serially
addressing a WL and a BL. A single selected device can be programmed by forcing a current through the BL with a voltage-controlled current source.
For reading a PCM cell, the selected BL is biased to a constant voltage of \unit[200]{mV}. The resulting read current is integrated by a capacitor,
and the resulting voltage is then digitized by the on-chip 8-bit cyclic analog-to-digital convertor (ADC). The total time of one read is
\unit[$1$]{$\mu$s}. The readout characteristic is calibrated via the use of on-chip reference poly-silicon resistors.

\subsection*{Generation of one million correlated processes and experimental details}
Let $\textbf{X}_r$ be a discrete binary process with probabilities $P(X_r(k) = 1) = p$ and $P(X_r(k) = 0) = 1-p$. Using $\textbf{X}_r$, $N$ binary
processes can be generated via the stochastic functions
\begin{eqnarray}\label{eqn:stochfun1}
P(X_i(k)=1|X_r(k) =1) &=& p + \sqrt{c}(1-p)\\\label{eqn:stochfun2} P(X_i(k)=1|X_r(k)=0) &=&p(1-\sqrt{c})\\\label{eqn:stochfun3} P(X_i(k)=0) &=& 1-
P(X_i(k)=1).
\end{eqnarray}
It can be shown that $E(X_i(k))=p$ and $Var(X_i(k))=p(1-p)$. Moreover, if two processes $\textbf{X}_i$ and $\textbf{X}_j$ are both generated using
Equations \ref{eqn:stochfun1} to \ref{eqn:stochfun3}, then the correlation coefficient between the two processes can be shown to be equal to $c$. For
the experiment presented, we chose an $\textbf{X}_r$ where $p=0.01$. A million binary processes were generated. Of these, $N_{c} = 95,525$ are
correlated with $c>0$. The remaining 904,475 processes are mutually uncorrelated. Each process is mapped to one pixel of a 1000 $\times$ 1000 pixel
black-and-white image of Albert Einstein: black pixels are mapped to the uncorrelated processes, and white pixels are mapped to the correlated
processes. The seemingly arbitrary choice of $N_c$ arises from the need to match with the white pixels of the image. The pixels turn on and off in
accordance with the binary values of the processes. One phase-change memory device is allocated to each of the one million processes.

\subsection*{Generation of weather data-based processes and experimental details}
The weather data was obtained from the National Oceanic and Atmospheric Administration (http://www.noaa.gov/) database of quality-controlled local
climatological data. It provides hourly summaries of climatological data from approximately 1600 weather stations in the United States of America.
The measurements were obtained over a 6-month period from January 2015 to June 2015 (181 days, 4344 hours). We generated one binary stochastic
process per weather station. If it rained in any given period of 1 hour in a particular geographical location corresponding to a weather station,
then the process takes the value 1; else it will be 0. For the experiments on correlation detection, we picked 270 weather stations with similar
rates of rainfall activity.

\section*{Acknowledgements}
We would like to thank the PCM teams at IBM Research - Zurich and the IBM T. J. Watson Research Center, USA. A.S. would like to acknowledge partial
financial support from the ERC Consolidator Grant 682675.
\section*{Author contributions}
A.S. and T.T. conceived the idea. A.S., N.P., M.L.G. and T.T. performed the experiments. A.S. and T.T. analyzed the data. L.K. and T.P. contributed
to the circuit design and signal-processing aspects, respectively. E.E. provided managerial support and critical comments. A.S. wrote the manuscript
with input from all the authors.
%\bibliographystyle{naturemag}
%\bibliography{papers_memcomputing}

\begin{thebibliography}{10}
\expandafter\ifx\csname url\endcsname\relax
  \def\url#1{\texttt{#1}}\fi
\expandafter\ifx\csname urlprefix\endcsname\relax\def\urlprefix{URL }\fi \providecommand{\bibinfo}[2]{#2} \providecommand{\eprint}[2][]{\url{#2}}

\bibitem{Y2016seshadriArXiV}
\bibinfo{author}{Seshadri, V.} \emph{et~al.}
\newblock \bibinfo{title}{Buddy-ram: Improving the performance and efficiency
  of bulk bitwise operations using dram}.
\newblock \emph{\bibinfo{journal}{arXiv preprint arXiv:1611.09988}}
  (\bibinfo{year}{2016}).

\bibitem{Y2013seshadriISM}
\bibinfo{author}{Seshadri, V.} \emph{et~al.}
\newblock \bibinfo{title}{Rowclone: fast and energy-efficient in-dram bulk data
  copy and initialization}.
\newblock In \emph{\bibinfo{booktitle}{Proceedings of the 46th Annual IEEE/ACM
  International Symposium on Microarchitecture}}, \bibinfo{pages}{185--197}
  (\bibinfo{organization}{ACM}, \bibinfo{year}{2013}).

\bibitem{Y2008strukovNature}
\bibinfo{author}{Strukov, D.~B.}, \bibinfo{author}{Snider, G.~S.},
  \bibinfo{author}{Stewart, D.~R.} \& \bibinfo{author}{Williams, R.~S.}
\newblock \bibinfo{title}{The missing memristor found}.
\newblock \emph{\bibinfo{journal}{Nature}} \textbf{\bibinfo{volume}{453}},
  \bibinfo{pages}{80--83} (\bibinfo{year}{2008}).

\bibitem{Y2011chuaAPA}
\bibinfo{author}{Chua, L.}
\newblock \bibinfo{title}{Resistance switching memories are memristors}.
\newblock \emph{\bibinfo{journal}{Applied Physics A}}
  \textbf{\bibinfo{volume}{102}}, \bibinfo{pages}{765--783}
  (\bibinfo{year}{2011}).

\bibitem{Y2015wongNatureNano}
\bibinfo{author}{Wong, H.-S.~P.} \& \bibinfo{author}{Salahuddin, S.}
\newblock \bibinfo{title}{Memory leads the way to better computing}.
\newblock \emph{\bibinfo{journal}{Nature Nanotechnology}}
  \textbf{\bibinfo{volume}{10}}, \bibinfo{pages}{191--194}
  (\bibinfo{year}{2015}).

\bibitem{Y2010borghettiNature}
\bibinfo{author}{Borghetti, J.} \emph{et~al.}
\newblock \bibinfo{title}{`memristive' switches enable `stateful' logic
  operations via material implication}.
\newblock \emph{\bibinfo{journal}{Nature}} \textbf{\bibinfo{volume}{464}},
  \bibinfo{pages}{873--876} (\bibinfo{year}{2010}).

\bibitem{Y2014kvatinskyTCAS}
\bibinfo{author}{Kvatinsky, S.} \emph{et~al.}
\newblock \bibinfo{title}{Magic: Memristor-aided logic}.
\newblock \emph{\bibinfo{journal}{IEEE Transactions on Circuits and Systems II:
  Express Briefs}} \textbf{\bibinfo{volume}{61}}, \bibinfo{pages}{895--899}
  (\bibinfo{year}{2014}).

\bibitem{Y2016zhaICCAD}
\bibinfo{author}{Zha, Y.} \& \bibinfo{author}{Li, J.}
\newblock \bibinfo{title}{Reconfigurable in-memory computing with resistive
  memory crossbar}.
\newblock In \emph{\bibinfo{booktitle}{Proceedings of the 35th International
  Conference on Computer-Aided Design}}, \bibinfo{pages}{120}
  (\bibinfo{organization}{ACM}, \bibinfo{year}{2016}).

\bibitem{Y2016vourkasIEEECSM}
\bibinfo{author}{Vourkas, I.} \& \bibinfo{author}{Sirakoulis, G.~C.}
\newblock \bibinfo{title}{Emerging memristor-based logic circuit design
  approaches: A review}.
\newblock \emph{\bibinfo{journal}{IEEE Circuits and Systems Magazine}}
  \textbf{\bibinfo{volume}{16}}, \bibinfo{pages}{15--30}
  (\bibinfo{year}{2016}).

\bibitem{Y2014ievlevNatPhys}
\bibinfo{author}{Ievlev, A.} \emph{et~al.}
\newblock \bibinfo{title}{Intermittency, quasiperiodicity and chaos in
  probe-induced ferroelectric domain switching}.
\newblock \emph{\bibinfo{journal}{Nature Physics}}
  \textbf{\bibinfo{volume}{10}}, \bibinfo{pages}{59--66}
  (\bibinfo{year}{2014}).

\bibitem{Y2013ielminiAM}
\bibinfo{author}{Cassinerio, M.}, \bibinfo{author}{Ciocchini, N.} \&
  \bibinfo{author}{Ielmini, D.}
\newblock \bibinfo{title}{Logic computation in phase change materials by
  threshold and memory switching}.
\newblock \emph{\bibinfo{journal}{Advanced Materials}}
  \textbf{\bibinfo{volume}{25}}, \bibinfo{pages}{5975--5980}
  (\bibinfo{year}{2013}).

\bibitem{Y2014lokePNAS}
\bibinfo{author}{Loke, D.} \emph{et~al.}
\newblock \bibinfo{title}{Ultrafast phase-change logic device driven by melting
  processes}.
\newblock \emph{\bibinfo{journal}{Proceedings of the National Academy of
  Sciences}} \textbf{\bibinfo{volume}{111}}, \bibinfo{pages}{13272--13277}
  (\bibinfo{year}{2014}).

\bibitem{Y2015burrITED}
\bibinfo{author}{Burr, G.~W.} \emph{et~al.}
\newblock \bibinfo{title}{Experimental demonstration and tolerancing of a
  large-scale neural network (165 000 synapses) using phase-change memory as
  the synaptic weight element}.
\newblock \emph{\bibinfo{journal}{IEEE Transactions on Electron Devices}}
  \textbf{\bibinfo{volume}{62}}, \bibinfo{pages}{3498--3507}
  (\bibinfo{year}{2015}).

\bibitem{Y2016shafieeISCA}
\bibinfo{author}{Shafiee, A.} \emph{et~al.}
\newblock \bibinfo{title}{Isaac: A convolutional neural network accelerator
  with in-situ analog arithmetic in crossbars}.
\newblock In \emph{\bibinfo{booktitle}{Proceedings of the 43rd International
  Symposium on Computer Architecture}}, \bibinfo{pages}{14--26}
  (\bibinfo{organization}{IEEE Press}, \bibinfo{year}{2016}).

\bibitem{Y2016chiISCA}
\bibinfo{author}{Chi, P.} \emph{et~al.}
\newblock \bibinfo{title}{Prime: A novel processing-in-memory architecture for
  neural network computation in {ReRAM}-based main memory}.
\newblock In \emph{\bibinfo{booktitle}{Proceedings of the 43rd International
  Symposium on Computer Architecture}}, \bibinfo{pages}{27--39}
  (\bibinfo{organization}{IEEE Press}, \bibinfo{year}{2016}).

\bibitem{Y2016bojnordiHPCA}
\bibinfo{author}{Bojnordi, M.~N.} \& \bibinfo{author}{Ipek, E.}
\newblock \bibinfo{title}{Memristive boltzmann machine: A hardware accelerator
  for combinatorial optimization and deep learning}.
\newblock In \emph{\bibinfo{booktitle}{IEEE International Symposium on High
  Performance Computer Architecture (HPCA)}}, \bibinfo{pages}{1--13}
  (\bibinfo{organization}{IEEE}, \bibinfo{year}{2016}).

\bibitem{Y2017burrAPX}
\bibinfo{author}{Burr, G.~W.} \emph{et~al.}
\newblock \bibinfo{title}{Neuromorphic computing using non-volatile memory}.
\newblock \emph{\bibinfo{journal}{Advances in Physics: X}}
  \textbf{\bibinfo{volume}{2}}, \bibinfo{pages}{89--124}
  (\bibinfo{year}{2017}).

\bibitem{Y2016burrJETCAS}
\bibinfo{author}{Burr, G.~W.} \emph{et~al.}
\newblock \bibinfo{title}{Recent progress in phase-change memory technology}.
\newblock \emph{\bibinfo{journal}{IEEE Journal on Emerging and Selected Topics
  in Circuits and Systems}} \textbf{\bibinfo{volume}{6}},
  \bibinfo{pages}{146--162} (\bibinfo{year}{2016}).

\bibitem{Y2011sebastianJAP}
\bibinfo{author}{Sebastian, A.}, \bibinfo{author}{Papandreou, N.},
  \bibinfo{author}{Pantazi, A.}, \bibinfo{author}{Pozidis, H.} \&
  \bibinfo{author}{Eleftheriou, E.}
\newblock \bibinfo{title}{Non-resistance-based cell-state metric for
  phase-change memory}.
\newblock \emph{\bibinfo{journal}{Journal of Applied Physics}}
  \textbf{\bibinfo{volume}{110}}, \bibinfo{pages}{084505}
  (\bibinfo{year}{2011}).

\bibitem{Y2015legalloNJP}
\bibinfo{author}{Le~Gallo, M.}, \bibinfo{author}{Kaes, M.},
  \bibinfo{author}{Sebastian, A.} \& \bibinfo{author}{Krebs, D.}
\newblock \bibinfo{title}{Subthreshold electrical transport in amorphous
  phase-change materials}.
\newblock \emph{\bibinfo{journal}{New Journal of Physics}}
  \textbf{\bibinfo{volume}{17}}, \bibinfo{pages}{093035}
  (\bibinfo{year}{2015}).

\bibitem{Y2014sebastianNatComm}
\bibinfo{author}{Sebastian, A.}, \bibinfo{author}{Le~Gallo, M.} \&
  \bibinfo{author}{Krebs, D.}
\newblock \bibinfo{title}{Crystal growth within a phase change memory cell}.
\newblock \emph{\bibinfo{journal}{Nature Communications}}
  \textbf{\bibinfo{volume}{5}}, \bibinfo{pages}{4314} (\bibinfo{year}{2014}).

\bibitem{Y2015ratyNatComm}
\bibinfo{author}{Raty, J.~Y.} \emph{et~al.}
\newblock \bibinfo{title}{Aging mechanisms in amorphous phase-change
  materials}.
\newblock \emph{\bibinfo{journal}{Nature Communications}}
  \textbf{\bibinfo{volume}{6}} (\bibinfo{year}{2015}).

\bibitem{Y2015corintoJETCAS}
\bibinfo{author}{Corinto, F.}, \bibinfo{author}{Civalleri, P.~P.} \&
  \bibinfo{author}{Chua, L.~O.}
\newblock \bibinfo{title}{A theoretical approach to memristor devices}.
\newblock \emph{\bibinfo{journal}{IEEE Journal on Emerging and Selected Topics
  in Circuits and Systems}} \textbf{\bibinfo{volume}{5}},
  \bibinfo{pages}{123--132} (\bibinfo{year}{2015}).

\bibitem{Y2016ascoliIJCTA}
\bibinfo{author}{Ascoli, A.}, \bibinfo{author}{Corinto, F.} \&
  \bibinfo{author}{Tetzlaff, R.}
\newblock \bibinfo{title}{Generalized boundary condition memristor model}.
\newblock \emph{\bibinfo{journal}{International Journal of Circuit Theory and
  Applications}} \textbf{\bibinfo{volume}{44}}, \bibinfo{pages}{60--84}
  (\bibinfo{year}{2016}).

\bibitem{Y2014lazerScience}
\bibinfo{author}{Lazer, D.}, \bibinfo{author}{Kennedy, R.},
  \bibinfo{author}{King, G.} \& \bibinfo{author}{Vespignani, A.}
\newblock \bibinfo{title}{The parable of google flu: traps in big data
  analysis}.
\newblock \emph{\bibinfo{journal}{Science}} \textbf{\bibinfo{volume}{343}},
  \bibinfo{pages}{1203--1205} (\bibinfo{year}{2014}).

\bibitem{Y2010liuCON}
\bibinfo{author}{Liu, S.-C.} \& \bibinfo{author}{Delbruck, T.}
\newblock \bibinfo{title}{Neuromorphic sensory systems}.
\newblock \emph{\bibinfo{journal}{Current opinion in neurobiology}}
  \textbf{\bibinfo{volume}{20}}, \bibinfo{pages}{288--295}
  (\bibinfo{year}{2010}).

\bibitem{Y2016tumaNatureNano}
\bibinfo{author}{Tuma, T.}, \bibinfo{author}{Pantazi, A.},
  \bibinfo{author}{Le~Gallo, M.}, \bibinfo{author}{Sebastian, A.} \&
  \bibinfo{author}{Eleftheriou, E.}
\newblock \bibinfo{title}{Stochastic phase-change neurons}.
\newblock \emph{\bibinfo{journal}{Nature Nanotechnology}}
  \textbf{\bibinfo{volume}{11}}, \bibinfo{pages}{693--699}
  (\bibinfo{year}{2016}).

\bibitem{Y2016legalloESSDERC}
\bibinfo{author}{Le~Gallo, M.}, \bibinfo{author}{Tuma, T.},
  \bibinfo{author}{Zipoli, F.}, \bibinfo{author}{Sebastian, A.} \&
  \bibinfo{author}{Eleftheriou, E.}
\newblock \bibinfo{title}{Inherent stochasticity in phase-change memory
  devices}.
\newblock In \emph{\bibinfo{booktitle}{46th European Solid-State Device
  Research Conference (ESSDERC)}}, \bibinfo{pages}{373--376}
  (\bibinfo{organization}{IEEE}, \bibinfo{year}{2016}).

\bibitem{Y2010closeIEDM}
\bibinfo{author}{Close, G.} \emph{et~al.}
\newblock \bibinfo{title}{Device, circuit and system-level analysis of noise in
  multi-bit phase-change memory}.
\newblock In \emph{\bibinfo{booktitle}{IEEE International Electron Devices
  Meeting (IEDM)}}, \bibinfo{pages}{29--5} (\bibinfo{organization}{IEEE},
  \bibinfo{year}{2010}).

\bibitem{Y2012pozidisIMW}
\bibinfo{author}{Pozidis, H.} \emph{et~al.}
\newblock \bibinfo{title}{A framework for reliability assessment in multilevel
  phase-change memory}.
\newblock In \emph{\bibinfo{booktitle}{4th IEEE International Memory Workshop
  (IMW)}}, \bibinfo{pages}{1--4} (\bibinfo{organization}{IEEE},
  \bibinfo{year}{2012}).

\bibitem{wikipedia}
\bibinfo{author}{Wikipedia}.
\newblock \bibinfo{title}{Albert {E}instein} (\bibinfo{year}{2016}).
\newblock \urlprefix\url{https://en.wikipedia.org/wiki/Albert_Einstein}.
\newblock \bibinfo{note}{Online; accessed: 2016-08-17}.

\bibitem{Y2015koelmansNatComm}
\bibinfo{author}{Koelmans, W.~W.} \emph{et~al.}
\newblock \bibinfo{title}{Projected phase-change memory devices}.
\newblock \emph{\bibinfo{journal}{Nature communications}}
  \textbf{\bibinfo{volume}{6}} (\bibinfo{year}{2015}).

\bibitem{Y2011xiongScience}
\bibinfo{author}{Xiong, F.}, \bibinfo{author}{Liao, A.~D.},
  \bibinfo{author}{Estrada, D.} \& \bibinfo{author}{Pop, E.}
\newblock \bibinfo{title}{Low-power switching of phase-change materials with
  carbon nanotube electrodes}.
\newblock \emph{\bibinfo{journal}{Science}} \textbf{\bibinfo{volume}{332}},
  \bibinfo{pages}{568--570} (\bibinfo{year}{2011}).

\bibitem{Y2012lokeScience}
\bibinfo{author}{Loke, D.} \emph{et~al.}
\newblock \bibinfo{title}{Breaking the speed limits of phase-change memory}.
\newblock \emph{\bibinfo{journal}{Science}} \textbf{\bibinfo{volume}{336}},
  \bibinfo{pages}{1566--1569} (\bibinfo{year}{2012}).

\bibitem{Y2003guetigJN}
\bibinfo{author}{G{\"u}tig, R.}, \bibinfo{author}{Aharonov, R.},
  \bibinfo{author}{Rotter, S.} \& \bibinfo{author}{Sompolinsky, H.}
\newblock \bibinfo{title}{Learning input correlations through nonlinear
  temporally asymmetric hebbian plasticity}.
\newblock \emph{\bibinfo{journal}{The Journal of neuroscience}}
  \textbf{\bibinfo{volume}{23}}, \bibinfo{pages}{3697--3714}
  (\bibinfo{year}{2003}).

\bibitem{Y2016tumaEDL}
\bibinfo{author}{Tuma, T.}, \bibinfo{author}{Le~Gallo, M.},
  \bibinfo{author}{Sebastian, A.} \& \bibinfo{author}{Eleftheriou, E.}
\newblock \bibinfo{title}{Detecting correlations using phase-change neurons and
  synapses}.
\newblock \emph{\bibinfo{journal}{IEEE Electron Device Letters}}
  \textbf{\bibinfo{volume}{37}}, \bibinfo{pages}{1238--1241}
  (\bibinfo{year}{2016}).

\bibitem{Y2011wrightAdvMat}
\bibinfo{author}{Wright, C.~D.}, \bibinfo{author}{Liu, Y.},
  \bibinfo{author}{Kohary, K.~I.}, \bibinfo{author}{Aziz, M.~M.} \&
  \bibinfo{author}{Hicken, R.~J.}
\newblock \bibinfo{title}{Arithmetic and biologically-inspired computing using
  phase-change materials}.
\newblock \emph{\bibinfo{journal}{Advanced Materials}}
  \textbf{\bibinfo{volume}{23}}, \bibinfo{pages}{3408--3413}
  (\bibinfo{year}{2011}).

\bibitem{Y2013wrightAFM}
\bibinfo{author}{Wright, C.~D.}, \bibinfo{author}{Hosseini, P.} \&
  \bibinfo{author}{Diosdado, J. A.~V.}
\newblock \bibinfo{title}{Beyond von-neumann computing with nanoscale
  phase-change memory devices}.
\newblock \emph{\bibinfo{journal}{Advanced Functional Materials}}
  \textbf{\bibinfo{volume}{23}}, \bibinfo{pages}{2248--2254}
  (\bibinfo{year}{2013}).

\bibitem{Y2015hosseiniEDL}
\bibinfo{author}{Hosseini, P.}, \bibinfo{author}{Sebastian, A.},
  \bibinfo{author}{Papandreou, N.}, \bibinfo{author}{Wright, C.~D.} \&
  \bibinfo{author}{Bhaskaran, H.}
\newblock \bibinfo{title}{Accumulation-based computing using phase-change
  memories with fet access devices}.
\newblock \emph{\bibinfo{journal}{IEEE Electron Device Letters}}
  \textbf{\bibinfo{volume}{36}}, \bibinfo{pages}{975--977}
  (\bibinfo{year}{2015}).

\bibitem{Y2011boniardiAPL}
\bibinfo{author}{Boniardi, M.} \& \bibinfo{author}{Ielmini, D.}
\newblock \bibinfo{title}{Physical origin of the resistance drift exponent in
  amorphous phase change materials}.
\newblock \emph{\bibinfo{journal}{Applied Physics Letters}}
  \textbf{\bibinfo{volume}{98}}, \bibinfo{pages}{243506}
  (\bibinfo{year}{2011}).

\bibitem{Y2012fantiniAPL}
\bibinfo{author}{Fantini, P.}, \bibinfo{author}{Ferro, M.},
  \bibinfo{author}{Calderoni, A.} \& \bibinfo{author}{Brazzelli, S.}
\newblock \bibinfo{title}{Disorder enhancement due to structural relaxation in
  amorphous ge2sb2te5}.
\newblock \emph{\bibinfo{journal}{Applied Physics Letters}}
  \textbf{\bibinfo{volume}{100}}, \bibinfo{pages}{213506}
  (\bibinfo{year}{2012}).

\bibitem{Y2016zipoliPRB}
\bibinfo{author}{Zipoli, F.}, \bibinfo{author}{Krebs, D.} \&
  \bibinfo{author}{Curioni, A.}
\newblock \bibinfo{title}{Structural origin of resistance drift in amorphous
  gete}.
\newblock \emph{\bibinfo{journal}{Physical Review B}}
  \textbf{\bibinfo{volume}{93}}, \bibinfo{pages}{115201}
  (\bibinfo{year}{2016}).

\bibitem{Y2015traversaTNNLS}
\bibinfo{author}{Traversa, F.~L.} \& \bibinfo{author}{Di~Ventra, M.}
\newblock \bibinfo{title}{Universal memcomputing machines}.
\newblock \emph{\bibinfo{journal}{IEEE Transactions on Neural Networks and
  Learning Systems}} \textbf{\bibinfo{volume}{26}}, \bibinfo{pages}{2702--2715}
  (\bibinfo{year}{2015}).

\bibitem{Y2015diventraSciAm}
\bibinfo{author}{Di~Ventra, M.} \& \bibinfo{author}{Pershin, Y.~V.}
\newblock \bibinfo{title}{Just add memory}.
\newblock \emph{\bibinfo{journal}{Scientific American}}
  \textbf{\bibinfo{volume}{312}}, \bibinfo{pages}{56--61}
  (\bibinfo{year}{2015}).

\bibitem{Y2015riosNaturePhotonics}
\bibinfo{author}{R{\'\i}os, C.} \emph{et~al.}
\newblock \bibinfo{title}{Integrated all-photonic non-volatile multi-level
  memory}.
\newblock \emph{\bibinfo{journal}{Nature Photonics}}
  \textbf{\bibinfo{volume}{9}}, \bibinfo{pages}{725--732}
  (\bibinfo{year}{2015}).

\bibitem{Y2007waserNatMat}
\bibinfo{author}{Waser, R.} \& \bibinfo{author}{Aono, M.}
\newblock \bibinfo{title}{Nanoionics-based resistive switching memories}.
\newblock \emph{\bibinfo{journal}{Nature Materials}}
  \textbf{\bibinfo{volume}{6}}, \bibinfo{pages}{833--840}
  (\bibinfo{year}{2007}).

\bibitem{Y2016kimIEDM}
\bibinfo{author}{Kim, W.} \emph{et~al.}
\newblock \bibinfo{title}{Ald-based confined pcm with a metallic liner toward
  unlimited endurance}.
\newblock In \emph{\bibinfo{booktitle}{IEEE International Electron Devices
  Meeting (IEDM)}}, \bibinfo{pages}{4--2} (\bibinfo{organization}{IEEE},
  \bibinfo{year}{2016}).

\bibitem{Y2007breitwischVLSI}
\bibinfo{author}{Breitwisch, M.} \emph{et~al.}
\newblock \bibinfo{title}{Novel lithography-independent pore phase change
  memory}.
\newblock In \emph{\bibinfo{booktitle}{IEEE Symposium on VLSI Technology}},
  \bibinfo{pages}{100--101} (\bibinfo{organization}{IEEE},
  \bibinfo{year}{2007}).

\end{thebibliography}

\end{document}